\documentstyle[preprint,aps,psfig]{revtex}
\tightenlines
   \newcommand{\be}{\begin{equation}}
   \newcommand{\ee}{\end{equation}}
   \newcommand{\bea}{\begin{eqnarray}}
   \newcommand{\eea}{\end{eqnarray}}

\begin{document}

\draft

\widetext
\title{Spin Dynamics near the Quantum Critical 
Point of Heavy-Fermions}
\author{C. P\'epin $^1$ and M. Lavagna $^2$}
\address{
$^1$ Department of Physics, MIT Ma 02139 Cambridge, US}
\address{
$^2$ Commissariat \`a l'Energie Atomique, DRFMC/SPSMS, 38054 Grenoble Cedex 9, France}

\maketitle \widetext
  \leftskip 10.8pt
  \rightskip 10.8pt
  \begin{abstract}
 The dynamical spin susceptibility is studied in the magnetically-disordered phase of heavy-Fermion systems near the antiferromagnetic quantum phase transition. In the framework of the $S=1/2$ Kondo lattice model, we introduce a perturbative expansion treating the spin and Kondo-like degrees of freedom on an equal footing. The general expression of the dynamical spin susceptibility that we derive presents a two-component behaviour:
a quasielastic peak as in a Fermi liquid theory, and a strongly q-dependent inelastic peak typical of a non-Fermi liquid behaviour. Very strikingly, the position of the inelastic peak is found to be pushed to zero at the antiferromagnetic transition with a vanishing relaxation rate. The comparison has been quantitatively made with Inelastic Neutron Scattering (INS) experiments
performed in $CeCu_{6}$ and  $Ce_{1-x}La_{x}Ru_{2}Si_{2}$. The excellent agreement that we have
found gives strong support to a two-band model with new prospects for the study of the quantum critical
phenomena in the vicinity of the magnetic phase transition.
  \par
  \end{abstract}
\vspace{0.2in}
\pacs{Keywords: Kondo lattice model - Dynamical spin susceptibility - Heavy Fermions}

\bigskip

One of the most striking properties of heavy Fermion compounds discovered
these last years is the existence of a quantum phase transition driven by
composition change (at $x_{C}=0.1$ in $CeCu_{6-x}Au_{x}$ and $x_{C}=0.08$ in
$Ce_{1-x}La_{x}Ru_{2}Si_{2}$), pressure or magnetic field. Important
insight is provided by the evolution of the low temperature neutron cross
section measured by Inelastic Neutron Scattering (INS) experiments when
getting closer to the magnetic instability. The experiments performed 
in pure compounds $CeCu_{6}$ and $CeRu_{2}Si_{2}$ by Regnault et al \cite{regnault}
and Aeppli et al \cite{aeppli} have shown the presence of two distinct contributions to
the dynamic magnetic structure factor: a $\overrightarrow{q}$-independent
quasielastic component, and a strongly $\overrightarrow{q}$-dependent
inelastic peak with a maximum at the value $\omega _{\max }$ of the
frequency. The former corresponds to localized excitations of Kondo-type
while the latter peaked at some wavevector ${\overrightarrow{Q}}$ is
believed to be associated with intersite magnetic correlations due to RKKY
interactions. Both the position and the width of
the inelatic peak vanish when getting near the magnetic
instability. 

Any theory aimed to describe the quantum critical phenomena in heavy-Fermion
compounds should account for the so-quoted behaviour of the dynamical spin
susceptibility. We start from the Kondo lattice model which is believed to
describe the physics of these systems. Most of the theories developed so far \cite{millis,auerbach86,lacroix79,rice} agree with
the existence of a hybridization gap which splits the Abrikosov-Suhl or
Kondo resonance formed at the Fermi level. The role of the interband
transitions has been outlined for long in order to explain the inelatic
component of the dynamical spin susceptibility. For instance, the theories
based on a 1/N expansion \cite{millis,auerbach86,lavagna,coleman,read} (where N is simultaneously the degeneracy of the
conduction electrons and of the spin channels) predict a maximum of $\chi
^{"}(k_{F},\omega )/ \omega$ at $\omega _{\max }$ of the order of the indirect
hybridization gap \cite{auerbach88}. However, the 1/N expansion theories present serious
drawbacks: (i) the spin fluctuation effects are automatically ruled out
since the RKKY interactions only occur at the following order in $1/N^{2}$ \cite{houghton},
(ii) they then fail to describe any magnetic instability and hence the
quantum critical phenomena mentioned above and (iii) the predictions for 
$\omega _{\max }$ and the associated relaxation rate cannot account for the
experimental observations near the magnetic instability. An improvement
brought by Doniach \cite{doniach87} and followed by other authors \cite{lacroix91} consists to consider the $1/N^{2}$ corrections in an instantaneous approximation: it gives back the
ladder diagram contribution to the dynamical spin susceptibility and then accounts
for the spin fluctuation effects. But still the
predictions for the frequency dependence of the dynamic magnetic structure factor
presents a gap of the order of the hybridization gap whatever the value of
the interaction is.  On the other hand, in front of the difficulties
encountered when starting from microscopic descriptions, various
phenomenological models (as the duality model of Kuramoto and Miyake \cite{kuramoto90} 
or that of Bernhoeft and Lonzarich \cite{bernhoeft}) have
been introduced to describe both the spin fluctuation and the
itinerant electron aspects with some successful predictions as the weak
antiferromagnetism of these systems.

In this work, we have developed a systematic approach to the Kondo lattice model
for $S=1/2$ ($N=2$) in which the Kondo-like and the spin degrees of freedom
are treated on an equal footing (see our longer paper \cite{pepin} for further details). The saddle-point results and the gaussian
fluctuations in the charge channel are consistent with the standard $1/N$
theories with the existence of two bands corresponding to the formation of a Kondo resonance splitted by a hybridization
gap.  For $n_c<1$, the chemical potential is located below the upper edge of the $\alpha$ band and the system is metallic. The gaussian fluctuations in the spin channel restore
the spin fluctuation effects which were missing in the $1/N$
expansion. 
The general expression of the dynamical spin
susceptibility $\chi _{ff}(\overrightarrow{q}, \omega)$ shows two contributions corresponding
to intra and interband transitions. Figure 1 reports the contiuum of intraband and interband particle-hole pair excitations. Far from the antiferromagnetic vector
$\overrightarrow{Q}$, $\chi _{ff}(\overrightarrow{q}, \omega)$ is dominated by the intraband contribution. In this case, the frequency dependence of the 
dynamical spin susceptibility is lorentzian in the low-frequency limit as in a standard Fermi liquid theory. On the other hand, we have shown how at the antiferromagnetic wave vector,  $\chi _{ff}({\overrightarrow{Q}}, \omega)$ is driven by the interband transitions giving rise to an inelastic peak. In a very striking way, 
the value $
\omega _{\max }$ of the maximum of $\chi _{inter}^{"}(\overrightarrow{Q},\omega )/{\omega}$
is pushed to zero at the antiferromagnetic phase transition and the inelastic
mode becomes soft with vanishing
relaxation rate. 

We believe that the two contibutions that we have found respectively correspond to the quasielastic and inelastic peaks oberved by INS. The role of the interband transitions had already been pointed out in
previous works \cite{auerbach88}. However while the previous studies conclude to the presence
of an inelastic peak at finite value of the frequency related to the
hybridization gap whatever the interaction J is, the originality of our approach is to predict a renormalization of both  $\omega _{\max }$ and the relaxation rate due to spin fluctuation effects. The comparison has been quantitatively made with Inelastic Neutron Scattering (INS) experiments performed in
$CeCu_{6}$ and  $Ce_{1-x}La_{x}Ru_{2}Si_{2}$. The quasielastic peak is typical of a Fermi liquid
while the other mode breaks the Fermi liquid description. For this reason
and because of the very good agreement of our results with the INS
experiments, our approach offers news prospects for the study of the quantum
critical phenomena in the vicinity of the antiferromagnetic phase transition.

\vfill\eject

\centerline {\bf FIGURE CAPTIONS}

\vspace{0.2in}

Figure 1: Continuum of the intra- and interband electron-hole pair excitations 
$\chi _{_{\alpha \alpha}}^{"}(q,\omega)\neq 0$ and $\chi _{_{\alpha \beta}}^{"}(q,\omega)\neq 0$.
Note the presence of a gap in the interband transitions equal to the indirect gap at $q=k_F$, and to the direct gap at $q=0$.
 
\vfill\eject

\vspace{2in}
\begin{figure}
\centerline{\psfig{file=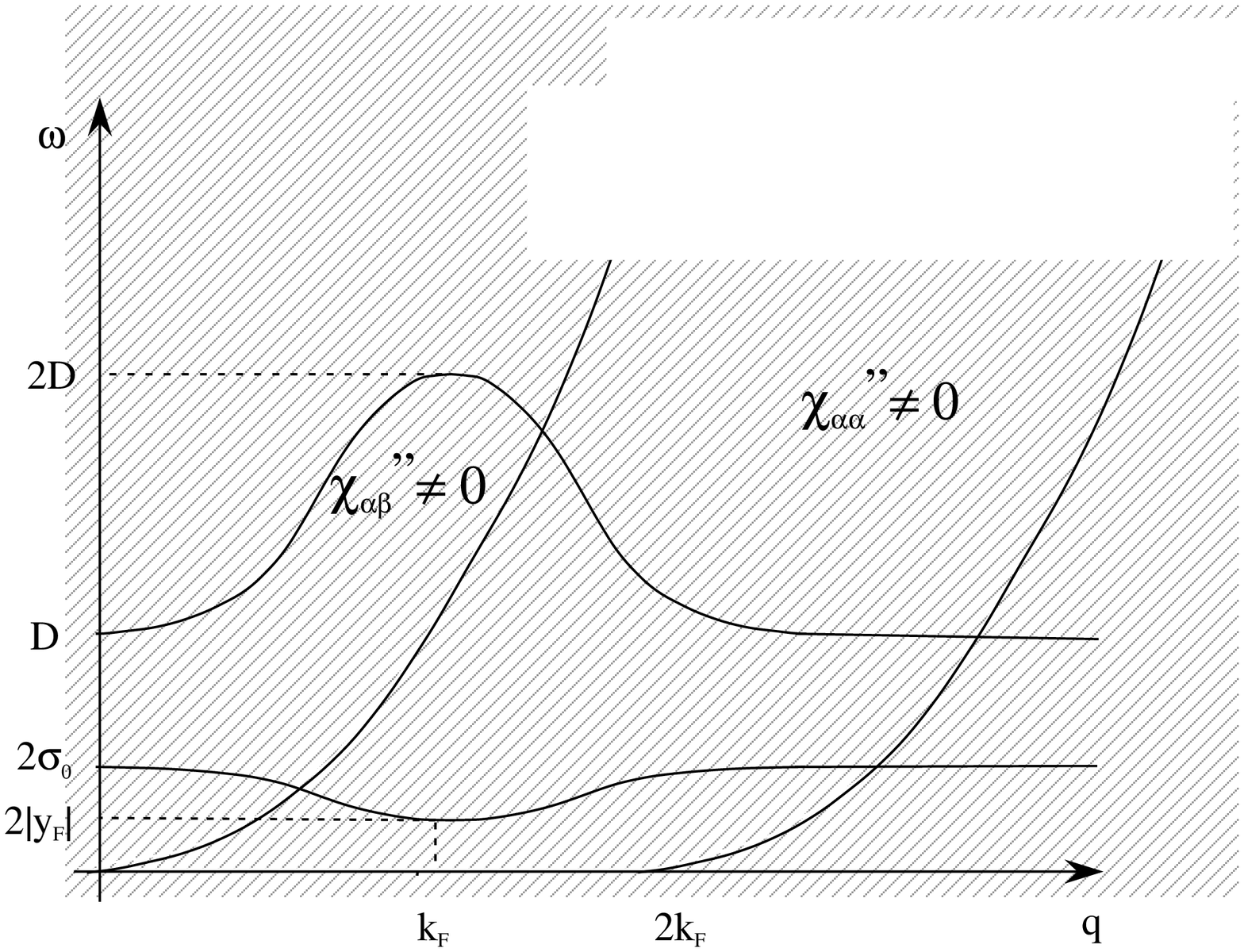,height=6cm,width=12cm}}
\vspace{2in}
\caption{}
\label{fig3}
\end{figure}

\vfill\eject


\bigskip


\begin{references}

\bibitem{regnault} L.P. Regnault, W.A.C. Erkelens, J. Rossat-Mignod, P. Lejay, 
J. Flouquet Phys. Rev. {\bf B 38}, 4481 (1988); S. Raymond, L.P. Regnault, 
S. Kambe, J.M. Mignod, P. Lejay, J. Flouquet J.Low Temp.Phys. {\bf 109}, 205 (1997) 
and this Proceedings

\bibitem{aeppli} G. Aeppli, H. Yoshizawa, Y. Endoh, E. Bucher, J. Hufnagl
Phys.Rev.Lett. {\bf 57}, 122 (1986)

\bibitem{millis} A.J. Millis, P.A. Lee Phys.Rev. {\bf B 35}, 3394 (1987)

\bibitem{auerbach86} A. Auerbach, K. Levin Phys.Rev.Lett. {\bf 57}, 877 (1986)

\bibitem{lavagna} M. Lavagna, A.J. Millis, P.A. Lee Phys.Rev.Lett. {\bf B 58},
266 (1987)

\bibitem{coleman} P. Coleman Phys.Rev. {\bf B 29}, 3035 (1984)

\bibitem{read} N. Read, D.N. Newns J.Phys.C {\bf 16}, 3273 (1983)

\bibitem{lacroix79} C. Lacroix, M. Cyrot Phys.Rev.{\bf B 20}, 1969 (1979)

\bibitem{rice} T.M. Rice and K. Ueda Phys.Rev.Lett. {\bf B 55}, 995 (1985)

\bibitem{auerbach88} A. Auerbach, Ju H. Kim, K. Levin, M.R. Norman Phys. Rev.
Lett. {\bf 60}, 623 (1988)

\bibitem{houghton} A. Houghton, N. Read, H. Won Phys.Rev. {\bf B 37}, 3782 (1988)  

\bibitem{doniach87} S. Doniach Phys.Rev. {\bf B 35}, 1814 (1987) 

\bibitem{lacroix91} C. Lacroix J.Magn.Magn.Mater. {\bf 100}, 90 (1991)

\bibitem{kuramoto90} Y. Kuramoto, K. Miyake J. Phys.Soc.Jpn {\bf 59}, 2831
(1990)

\bibitem{bernhoeft} N.R. Bernhoeft, G.G. Lonzarich J.Phys.Cond.Mat. {\bf 7},
7325 (1995) 

\bibitem{pepin} C. P\'epin, M Lavagna Cond-Mat Preprint 9803255

\end{references}
\end{document}